\documentclass[pra,aps,superscriptaddress]{revtex4}
\usepackage{amsmath}
\usepackage{amsfonts}
\usepackage{amssymb}
\usepackage[dvips]{graphicx}

\usepackage{txfonts}
\usepackage{psfrag}
\usepackage{epsfig}
\usepackage{cancel}
\usepackage[all]{xy}
\usepackage{ulem}

\newcommand\ii{\leavevmode\hbox{\small1\kern-3.3pt\normalsize1}}

\newcommand{\bra}{\langle}
\newcommand{\ket}{\rangle}
\newcommand{\al}{\vec{\alpha}}
\newcommand{\be}{\vec{\beta}}
\newcommand{\tet}{\vartheta}
\newcommand{\fii}{\varphi}
\newcommand{\tr}{\hbox{Tr}}
\newcommand{\alphas}{{\al_1, \dots ,\al_N}}
\newcommand{\mus}{{\mu_1, \dots ,\mu_N}}
\newcommand{\sqd}{\frac{1}{\sqrt{2}}}
\newcommand{\cleb}[6]{
    C\left.\!\!\!{ \begin{array}{lll}
    \scriptstyle{#1}  &\!\!\! \scriptstyle{#2} &\!\!\! \scriptstyle{#3} \\
    \scriptstyle{#4}  &\!\!\! \scriptstyle{#5} &\!\!\! \scriptstyle{#6}
    \end{array}}\right.
}

\begin{document}

\title{Entanglement detection with bounded reference frames}

\author{Fabio Costa}
\affiliation{Institute for Quantum Optics and Quantum Information,
Austrian Academy of Sciences, Boltzmanngasse 3, A-1090 Vienna,
Austria} \affiliation{Faculty of Physics, University of Vienna,
Boltzmanngasse 5, A-1090 Vienna, Austria}

\author{Nicholas Harrigan}
\affiliation{Department of Physics, Imperial College London, Prince Consort Rd, London SW7 2BW}

\author{Terry Rudolph}
\affiliation{Department of Physics, Imperial College London, Prince Consort Rd, London SW7 2BW}
\affiliation{Institute for Mathematical Sciences, 53 Princes Gate, Exhibition Rd, London SW7 2PG}

\author{{\v C}aslav Brukner}
\affiliation{Institute for Quantum Optics and Quantum Information,
Austrian Academy of Sciences, Boltzmanngasse 3, A-1090 Vienna,
Austria} \affiliation{Faculty of Physics, University of Vienna,
Boltzmanngasse 5, A-1090 Vienna, Austria}

\date{\today}

\begin{abstract}
Quantum experiments usually assume the existence of perfect,
classical, reference frames, which allow for the specification of measurement
settings (e.g. orientation of the Stern Gerlach magnet in spin
measurements) with arbitrary precision. If the reference frames are
``bounded'' (i.e. quantum systems themselves, having a finite number of degrees of freedom), only limited
precision can be attained. Using spin coherent states as bounded
reference frames we have found their minimal size needed to violate local
realism for entangled spin systems. For composite systems of
spin-$1/2$ particles  reference frames of very small size are
sufficient for the violation; however, to see this violation for
macroscopic entangled spins, the size of the reference frame must be at least
quadratically larger than that of the spins. Unavailability of such reference frames gives a possible
explanation for the non-observance of violation of local realism in everyday experience.
\end{abstract}


\maketitle

The Kochen-Specker test of contextuality~\cite{ks} and Bell's test of
local realism~\cite{bell} provide theory-independent tests of the
``classicality'' of a system. In the latter, correlations between
space-like separated parts of a composite system are measured for
different choices of measurement settings. Certain combinations of
these correlations constitute ``test quantities'' (Bell's
inequalities) that are bounded in all classical (local realistic) theories.
Violation of these bounds imply that the tested state has no local
realistic explanation. Typically, the choices of the settings correspond to
different orientations of the measuring apparatuses (like a
polarizer or the Stern-Gerlach magnet). This implicitly assumes the
existence of an external (classical)  reference frame (RF), which allows one to specify with
arbitrary precision the directions chosen. But, what if no RF is
available? The impossibility of specifying measurement settings
precisely, in the absence of a perfect RF,  leads to a kind of ``intrinsic decoherence'' \cite{bartlett, poulin, gambini} that might
wash out all quantum features. \textit{What are the minimal RF resources
such that quantum features of a given system can still be observed?}

If we adopt the natural assumption that physical resources in the
universe are finite, we will always be confronted with bounded RFs.
Physically, this means that our measurements will
always be imprecise. It is of fundamental interest to determine
the minimal measurement precision required such that one can still observe
genuine quantum features, such as contextuality or violation of local
realism (see, for example, ~\cite{peres}, ~\cite{Kofl2006}). In addition to these foundational
reasons, the questions given above are relevant for developing
methods to contend with bounded RFs using relational encoding,
particularly in the context of computation, cryptography, and
communication~\cite{refrev, refbits, info1, info2, orbital1, orbital2}. For certain tasks, such as quantum key
distribution~\cite{QKD} or quantum communication
complexity~\cite{CCP}, entangled states are useful only to the
extent that they violate Bell's inequalities. It is thus important
to quantify the costs in RF resources for the violation.

To introduce the idea of measuring relative degrees of freedom in
the situation of lacking an external RF consider the problem of
determining the direction towards which a spin-1/2 particle points. In
general the direction can be defined as a relative angle to some
macroscopic pointer (e.g. the Stern-Gerlach magnet), which serves as
an external RF. Performing many repetitions of the Stern-Gerlach
experiment with the same spin state, we can infer this angle. If we
are given two spin-1/2 particles, one can determine the relative
angle between them, by first measuring the angles between each of
the spins and the external RF and from these computing the relative
angle. Now suppose that the experimenter has no access to such an external Cartesian RF; operationally, this means that she has no information about her orientation with respect to the rest of the world  (but she can still control all the devices in her laboratory). In particular, the angles between her instruments and those used to prepare the particles are not known and may change in every repetition of the experiment.
Nonetheless, the relative orientation between the two spins can be
measured in a manner which is invariant under rotations. If she
measures the total spin of the two particles, this can take the values 0
and 1. Now we are tempted to say that the two spins add when
they are aligned and subtract when they are anti-aligned, and so we can
interpret this as the measurement of the projection of the first
spin along the direction of the second.
This procedure leads unavoidably to errors: e.g. if the
spins were initially in the state
$|\psi^+\ket=\frac{1}{\sqrt{2}}(|z+\ket|z-\ket+|z-\ket|z+\ket)$, then a measurement on each particle with respect to an external RF along $z$ would imply that they are anti-aligned along this direction. However, the measurement of total spin is interpreted as if they were aligned, because the total spin of $|\psi^+\ket$ is 1.  As this procedure is proven to be optimal
~\cite{optimal}, there is a
fundamental restriction in the determination of the relative angle
between two (finite) spins that the experimenter can achieve in absence of an external reference frame. One can apply the same procedure using a spin-$j$ coherent state as the RF; as $j$ becomes larger, the errors introduced decrease and, eventually, for $j\rightarrow \infty$ (unbounded RF) the measurement with the classical RF is exactly reproduced \cite{note}.

In this work we determine how ``strong'' RFs need to be to allow
violation of Bell's inequalities. In the Bell test the observers are
given bounded RFs . Since we are interested in violations
of local realism in the transition from quantum to classical RFs, we
use the spin coherent states to represent quantum RFs as they are
closest to the notion of a classical direction. We find that a pair of spin-1/2 particles
exhibits violation of Clauser-Horne-Shimony-Holt (CHSH)
inequality already for $j_{RF} \!> \! \frac{5}{2}$. In the case of
multi-particle Mermin~\cite{mermin} inequalities, even if half of
the RFs are of minimal size $j_{RF}\!=\!1/2$, and the other half
``unbounded'' (classical), the ratio between the quantum and local
realistic bound remains exponential in the number of entangled
spins. Finally, for the
case of two macroscopic spins exhibiting violations of a Bell
inequality when classical RFs are available, we find that the
violation  is possible with bounded RFs only if their size is {\it
quadratically} larger than the size of the spins. Since our everyday
RFs do not meet this requirement of macroscopically large spins,
this suggests an explanation as to why we do not see such violations
in everyday life. All our results are derived for violations of
local realism, but - when applied to different degrees of freedom of a single system - they can can be interpreted as
contextuality proofs as well \cite{rauch}.


\section{Measurement of relative degrees of freedom.} In this work we
consider directional RFs. Given a classical, external RF, one can
measure the projection of a spin-$j_{S}$ particle along a specific
direction, where the $2j_{S}+1$ possible outcomes
$m=-j_{S},\dots,j_{S}$, correspond to the projectors $\Pi_m =
|j_{S},m\ket\bra j_{S},m|$. Generalizing this we consider bounded
RFs by replacing the classical RF by a quantum RF in the form
of a coherent state of spin $j_{RF}$.

Without an external RF, the experimenter can only measure relational
degrees of freedom of the particle and the coherent state $|j_{RF}\ket$. The task
is thus to estimate the relative angle between them using only rotationally invariant operations. An optimal procedure consists in performing the projective
measurement onto the subspaces of total spin $j$ of two spins~\cite{refrev},  which
can take the values $j = |j_{RF}-j_{S}|,\dots,j_{RF}+j_{S}$.  When
the outcome $j = j_{RF} + m$ of  total spin is observed, we
associate the spin component $m$ along the direction of the bounded RF
to the system. In this way the spin projection measurement relative
to a bounded RF simulates the one relative to an unbounded RF. The
projectors associated to subspaces of total spin are
\begin{displaymath}
    \hat{\Pi}_{j_{RF}+m} = \sum^{j_{RF}+m}_{m'=-j_{RF} - m}
|j_{RF}+m,m'\ket\bra j_{RF}+m,m'|\;.
\end{displaymath}

The effective measurement on the system alone is represented by the POVM elements
\begin{equation} \label{povm}
    \hat{P}^{j_{RF}}_{j_{S} \, m} = \bra j_{RF} | \hat{\Pi}_{j_{RF}+m} |j_{RF} \ket  \; .
\end{equation}
These can be expressed in terms of
the Clebsch-Gordan coefficients $\cleb{j_1}{j_2}{j}{m_1}{m_2}{m} :=
\bra j,m||j_1,m_1\ket|j_2,m_2\ket$ and are given
by $\hat{P}^{j_{RF}}_{j_S \, m} = \sum_{m_S n_S}
\cleb{j_{RF}}{j_S}{j_{RF}+m}{j_{RF}}{m_S}{j_{RF}+m_S}
\cleb{j_{RF}}{j_S}{j_{RF}+m}{j_{RF}}{n_S}{j_{RF}+n_S}|j_S,m_S\ket\bra
j_S,n_S|$. Using the asymptotic properties of the Clebsch-Gordan
coefficients~\cite{mat} it can be shown that $ \hat{P}^{j_{RF}}_{j_S
\, m} \rightarrow |j_S,m\ket\bra j_S,m| \equiv \hat{\Pi}_m$ when
$j_{RF}\rightarrow \infty $. This shows that the relational measurement,
with increasingly larger bounded RF, tends to
the measurement with unbounded RF.

A more general way to exploit relative degrees of freedom would be to encode information in rotationally invariant subsystems. For example, three spin-$\frac{1}{2}$ particles possess a two-level invariant subsystem in the subspace of total spin $\frac{1}{2}$. Thus, one rotationally invariant qubit (quantum bit) can be encoded in three physical ones and six spin-$\frac{1}{2}$ particles (three for Alice and three for Bob) would allow perfect violation of Bell's inequalities with no need of a rotational RF. In this work we do not consider such general schemes. Rather, we study the situations in which "system" and "quantum reference frame" are separated - resembling the conventional situation in experiments in which "system" and "classical reference frame" are separated. This will allow us to investigate quantum features of systems as the bounded quantum reference frames approach the classical limit.

In the Bell experiment each of the observers chooses between two or
more measurement settings, corresponding, for example, to
measurements of spin components along different directions $\al_1$,
$\al_2$,... In our scenario this choice corresponds to the use of
coherent states $|\al_1\ket$, $|\al_2\ket$,... pointing to different
directions as RFs. Here $|\al_i\ket$ is the eigenstate with the
maximal eigenvalue of the spin component along the direction
$\al_i$. One possible way to prepare such states is to apply an
appropriate rotation to the given coherent state $|j_{RF}\rangle$
pointing towards $z$ axis. Since, however, Alice and Bob are assumed
to have no RF, we consider the following operational
realization. In every experimental run a third party (Charlie) sends
to both Alice and Bob one coherent state for each
possible setting, together with the entangled pair to be measured.
Each coherent state is prepared along the direction which would be
chosen when classical RF were used. Having no RF for directions,
Alice cannot know the angles of the different coherent states, but
she can still distinguish between them (for example, Charlie can
send them with a short, agreed, time delay), then Alice can decide
which one to use as a RF, allowing the freedom of choice necessary
in a Bell experiment.

In an alternative implementation, Charlie chooses the setting and sends only the corresponding (one) coherent state. In this case no additional resources than those actually used are distributed, but the freedom of choice of the two observers is not strictly satisfied. One could think of more complicated schemes for ensuring that (e.g., introducing more agents that share a global RF with Charlie, but are space-like separated, such that they can perform the settings choice under locality conditions), however, in the present work the focus is not on a strict disproof of local realistic theories, but rather on the ability to measure quantum correlations under the restriction of bounded RF.

It is here assumed that the channel between
Charlie and the observers is subject to a collective noise, that is
to say, all the particles sent to one observer undergo the same
unknown rotation (but a different rotation occurs in different runs
and for the different observers), this is important since we want
to exploit relative degrees of freedom. Such an assumption could be
reasonable in some quantum communication schemes.


\section{Violation of Bell's inequalities.} As a first example we
consider a Bell experiment on an entangled pair of spin-$1/2$
particles with bounded RFs. We will determine the minimal size for the
spin RFs such that the outcomes can still violate Bell's inequalities.

We consider the CHSH inequality ~\cite{chsh}
\begin{equation} \label{chsh}
  \hspace{-1 pt}  S:=\left|E(\al_1,\be_1)+E(\al_1,\be_2)+E(\al_2,\be_1)-E(\al_2,\be_2)\right|\; \leq
    2,
\end{equation}
where $E(\al_i,\be_j)$ is the correlation function for the
measurement  $\al_i$ at one laboratory and $\be_j$ at the other
laboratory. In quantum mechanics, for a given state $\hat{\rho}$ of
the pair, if the first spin is measured along direction $\al_i$ and
the second along $\be_j$, the correlation function reads
$E(\al_i\,,\:\be_j)=\tr\left[\hat{\rho}\: \left(\al_i\cdot
\vec{\sigma}\right)\:\left(\be_j\cdot\vec{\sigma}\right)\right]$.

In contrast to the standard Bell experiment in which two
distant observers possess unbounded RFs, we assume that they can only
use their coherent states ($|j_1\ket$ and $|j_2\ket$) with respect
to which entangled spins can be measured.
How large must $j_1$ and $j_2$ be, such that the CHSH inequality is still violated?

We assume that the pair is in the singlet state $|\psi^-\ket = \frac{1}{\sqrt{2}} \left(|z+\ket|z-\ket-|z-\ket|z+\ket\right)$.
The two observers can choose between two measurement settings each, the setting being defined by the direction towards which the RF coherent state is pointing.
As $|\psi^-\ket$ is a rotationally invariant state, the only relevant
parameter in the correlation function is the relative angle between
the two pointers. It is more mathematically convenient (but operationally equivalent) to write the state with
fixed measurement settings and then to apply a rotation of an angle $\tet$ to one of the particles.
The corresponding rotated singlet state
is $|\psi^-(\tet)\ket =\frac{1}{\sqrt{2}}
\left[\sin\frac{\tet}{2}\left( |z+\ket|z+\ket + |z-\ket|z-\ket
\right)+ \cos\frac{\tet}{2} \left(|z+\ket|z-\ket - |z-\ket|z+\ket
\right) \right]$.
If one measures total spin of the joint system of particle and RF at
the two laboratories, the probabilities of the various outcomes are

\begin{displaymath}
p_{m\, n}(\tet) =\sum_{m_1 \, m_2}\hspace{-0.1 cm}\big| \bra
j_1|\bra j_2|\bra \psi(\tet)| |j_1\! + m ,j_1\! + m_1 \ket |j_2\! +
n , j_1\! + m_2 \ket \big|^2\,,
\end{displaymath}
where $m , n =
-\frac{1}{2},\frac{1}{2}$. (i.e.  $p_{\frac{1}{2}\, \frac{1}{2}}(\tet)$ is
the probability of finding the two particles aligned along
$|j_1\ket$ and $|j_2\ket$ respectively,
$p_{-\frac{1}{2}\, -\frac{1}{2}}(\tet)$ is the probability of
finding both anti-aligned, etc). To calculate these one needs the coefficients
\begin{eqnarray} \label{clebsch}
    && \bra j|\bra z + | |j + m,j+ n\ket = \delta_{n,\frac{1}{2}}\delta_{m,\frac{1}{2}}\; ,\cr
    && \bra j|\bra z - | |j + m, j+ n\ket = \frac{\delta_{n,\:-\frac{1}{2}}}{\sqrt{2j+1}}\left(\delta_{m,\frac{1}{2}}+\sqrt{2j}\delta_{m,-\frac{1}{2}}\right) .
\end{eqnarray}
Inserting the probabilities in the definition of the correlation
function $E^{j_1 j_2}(\tet) = -\sum_{m \, n =
-\frac{1}{2}}^{\frac{1}{2}}(-1)^{m+n}p_{m\,n}(\tet)$ one obtains
\begin{equation}
    E^{j_1 j_2}(\tet) = \frac{1-4j_1j_2\cos\tet}{(2j_1+1)(2j_2+1)}\;.
    \label{corr1}
\end{equation}

In the limit of large $j_1$ and $j_2$, Eq.~\eqref{corr1} becomes the
familiar expression $E(\tet)=-\cos\tet$ for the singlet correlation
function with unbounded RFs.
Note that, differently from
  the case with classical RFs, we have in \eqref{corr1} an offset term before $\cos\tet$, such that $E^{j_1 j_2}(\tet) \neq E^{j_1 j_2}(\tet+\pi)$. This
  implies that the measurement settings which maximize
\eqref{chsh} are for the relative angle
$\frac{3}{4}\pi$ in all cases except between $\al_2$ and $\be_2$, for
which the angle is $\frac{\pi}{4}$.
The RF-dependent CHSH
expression then reads
\begin{equation}
        S(j_1,\,j_2) = 2\left|\frac{1+4\sqrt{2}j_1j_2}{(2j_1+1)(2j_2+1)} \right|\; .
\end{equation}
It exceeds the local realistic limit of 2 if $j_1 > \frac{j_2}{2(\sqrt{2}-1)j_2-1}$. Therefore, for equal RFs, one thus needs at least
$j_1=j_2=\frac{5}{2}$.


\section{Mermin inequalities.} We explore violation of multi-particle
Bell's inequalities with bounded RFs. Consider $N$ spin-$1/2$
particles (systems $S_1,..., S_N$), that are measured along
directions $\alphas$. Each individual measurement can give $\pm
\frac{1}{2}$ as result; a specific outcome is thus labeled by a
string $\{\mus\}$, where $\mu_k\!=\!1$ stands for the $k$-th spin
detected aligned with $\al_k$, while $\mu_k\!=\!-1$ represents the
spin anti-aligned with $\al_k$. The multiparticle correlation
function is defined as $
    \!\!E(\alphas)=\!\!\sum_{\mus=\pm 1} \prod^{N}_{k=1}\mu_k  p(\mus\,;\,\alphas),
$
where $p(\mus\,;\,\alphas)$ is the probability for the outcomes $\mus$ given the settings $\alphas$.

The Mermin inequality is given by \cite{mermin, marek}
\begin{equation}
        \label{mermin}
          M := \Big|\!\!\!\!\!\!\!\! \sum_{x_1, \dots\, ,x_N=0,1}\hspace{-0.4 cm} \cos[\frac{\pi}{2}(x_1+\dots +x_N)]
          E(\al_{x_1}, \dots\, ,\al_{x_N}) \Big| \leq 2^{\frac{N-1}{2}}.
\end{equation}
Using unbounded RFs the Mermin expression reaches its maximal value
of $M=2^{N-1}$ for the Greenberger-Horne-Zeilinger (GHZ) state $|\psi\ket =   \sqd \left( \otimes^{N}_{k=1}|z+\ket_{S_k} +
\otimes^{N}_{k=1}|z-\ket_{S_k} \right)$ and measurement settings $\al_0=X\equiv(\frac{\pi}{2},0)$ and
$\al_1=Y\equiv(\frac{\pi}{2},\frac{\pi}{2})$ for every particle.
Note that the ratio between maximal quantum and local realistic
bound increases exponentially with the number of particles: $
2^{\frac{N-1}{2}}$.

Again, we assume now that the $k$-th observer, $k=1,...,N$, is given a bounded
RF in form of the coherent state $|j_k\ket$. Each of the observers measures along directions $\al_0 \equiv
(\tet_0,\fii_0)$ and $\al_1 \equiv (\tet_1,\fii_1)$. Rewriting the particles in the GHZ state in terms of these directions, we have $|\psi(\alphas)\ket =\sqd\left(\otimes^{N}_{k=1}|\al_k+\ket_{S_k} +\otimes^{N}_{k=1}|\al_k-\ket_{S_k} \right)$, where $|\al_k\pm\ket$ is the state of a particle after the inverse rotation along $\al_k$ is applied to $|z\pm\ket$.

As for the two-particle case, in the $k$-th laboratory the total spin of the joint system $k$-th
particle + $k$-th RF is measured and the outcome is interpreted as the projection of the particle's spin along the RF's direction.
After a somewhat lengthy but straightforward calculation one obtains the correlation
function observed to be:
\begin{eqnarray}
\label{corrmer}
    && E(\alphas;j_1,\dots, j_N)=\frac{1}{\prod^{N}_{k=1}d_k}
    \left\{ \frac{1}{2}\left[\prod^{N}_{k=1}(1+2j_k\cos\tet_k) 
    + \prod^{N}_{k=1}(1-2j_k\cos\tet_k)\right] \right. \cr
    &&  \;\;\;\;\;\;\;\;\;\;\;\;\;\;\;\;\;\;\;\;\;\;\;\;    \left. + \cos(\sum^{N}_{k=1}\fii_k)\prod^{N}_{k=1}2j_k\sin\tet_k \right\} ,
\end{eqnarray}
where $d_k = 2j_k+1$.

Inserting the correlation function~(\ref{corrmer}) into the
left-hand side of the inequality~\eqref{mermin} we find for the Mermin
expression:
\begin{equation}  \label{mermin1}
        \hspace{-0.88 cm}   M(j_1,\dots,j_k)  = \frac{1}{\prod^{N}_{k=1}d_k}\left|\sqrt{2}\cos(N \frac{\pi}{4})+ 2^{N-1}\prod^{N}_{k=1}(2j_k) \right| .
\end{equation}
For $j_k\rightarrow \infty$, this approaches the value $2^{N-1}$
when unbounded RFs are used.
In the limit of large number of particles, the Mermin expression becomes
\begin{equation}
        \label{largemermin}
        M = 2^{N-1}\prod^{N}_{k=1}\frac{j_k}{j_k+\frac{1}{2}} + {\cal O}(1)\,,\; N\rightarrow\infty.
\end{equation}
If all the RFs are of the same size, $j_k=j$  $\forall\, k$, the
minimal size of RFs that leads to violation is $j=3/2$. One can use,
however, even fewer resources if one allows spins of different
lengths for RFs.
If one takes $N_1$ spins of size $j_1$ and $N_2$ of size $j_2$,
expression (\ref{largemermin}) becomes $M \approx
2^{N-1}\left(\frac{j_1}{j_1+\frac{1}{2}}
\right)^{N_1}\left(\frac{j_2}{j_2+\frac{1}{2}}  \right)^{N_2}$. In this case the
Mermin inequality is violated if
\begin{equation} \label{merminlarge}
        \left(\sqrt{2} \frac{j_1}{j_1+\frac{1}{2}}  \right)^{N_1}\left(\sqrt{2} \frac{j_2}{j_2+\frac{1}{2}} \right)^{N_2} > \sqrt{2} .
\end{equation}
As both factors are positive, \eqref{merminlarge} can hold only if at least one of the two is larger
than 1, which is equivalent to $j_i>\frac{1}{2(\sqrt{2}-1)}=1.21$. This
implies that some of the RFs must have spin size equal to or larger
than $\frac{3}{2}$. If the parties  have $N_1$ spin-$\frac{3}{2}$ RFs and
$N_2=N-N_1$ spin-$\frac{1}{2}$ RFs, the minimal ratio is
$\frac{N_1}{N} \cong 0.85$ for seeing violation. Therefore the minimal resources needed is 85\% of spin-$\frac{3}{2}$ and 15\% of spin-$\frac{1}{2}$ reference frames.

Another interesting case is when a fraction of the RFs is
unbounded, which is equivalent to taking the limit
$j_2\rightarrow\infty$ in the inequality ~\eqref{merminlarge}. For $N_1$ RFs of
size $j_1$ and $N_2$ unbounded RFs, it becomes $\left(\sqrt{2} \frac{j_1}{j_1+\frac{1}{2}}  \right)^{N_1} >
2^{\frac{1-N_2}{2} }$. 
For $j_1=1/2$, this is satisfied when $N_2>N_1+1$, which
means that when half of the RFs are unbounded and half are
as small as spin-$\frac{1}{2}$, violation of the Mermin
inequality is still possible.

Note that in all cases considered -- even when using small quantum
RFs -- the ratio between the quantum and local realistic bound is
still exponential, as can be easily seen by inserting the results
found into the expression~\eqref{largemermin}. However, if a single
measurement is replaced with a random guess (corresponding
mathematically to $j_1=0$), the inequality is satisfied. Thus a non-trivial RF is required for every observer in order to see nonclassicality.

\section{Higher spins and the classical limit.} It was shown in ~\cite{peres} that
violation of Bell's inequalities with entangled systems of arbitrarily large dimension is
possible. This shows that the view that large quantum numbers are associated with the classical limit is, in general, erroneous. We will show that
to observe violations of local realism for large spins it is necessary to use
the RFs of size sufficiently large compared to the size of the
spins. The scaling of the two sizes is the issue we are interested in.

Following Peres~\cite{peres} we consider a pair of spin-$j_S$
particles in the generalized singlet state:
\begin{equation}
        \label{singlet}
        |\Psi_{j_S}^-\ket := \frac{1}{\sqrt{2j_S+1}}\sum^{j_S}_{m=-j_S}(-1)^{j_S-m}|j_S,m\ket|j_S,-m\ket .
\end{equation}
and define the parity measurement

\begin{equation}
    \hat{P}^{\,c} = \sum^{j_S}_{m=-j_S}(-1)^{j_S-m}\hat{\Pi}_m \;,
    \label{parity}
\end{equation}
with $\hat{\Pi}_m = |j_s,m\ket\bra j_s,m|$ the projectors onto
subspaces of the spin component along the $z$ axis. The parity
measurement takes the value $+1$ for all even $m$ , and $-1$ for all
odd $m$. When the parity operator is defined with respect to spin
projection along some other direction $\al$, we will speak about
parity measurement $P^{\,c}(\al)$ along this direction. For Alice's
measurement along the direction $\al$ and Bob's along $\be$, the
correlation function is defined as
$E(\al,\,\be)=\bra\Psi_{j_S}^-|\hat{P}^{\,c}(\al)\otimes
\hat{P}^{\,c}(\be)|\Psi_{j_S}^-\ket$. The CHSH
inequality~\eqref{chsh} is violated for parity measurements in the
singlet state for arbitrarily large spins~\cite{peres, garg}.

To consider violation of the inequality with bounded RFs we
introduce a coherent state of length $j_{RF}$ for each observer,
and replace the projectors $\hat{\Pi}_m$ in \eqref{parity} with the POVM from
Eq.~\eqref{povm}.
When the measurement setting $\al =
(\tet,\,\fii)$ is chosen, the coherent state $|\al \ket$ aligned in that direction is used.
In the basis of the spin projection along $z$-axis, it reads $|\al \ket \equiv |\tet,\,\fii\ket = \sum^{j_{RF}}_{m=-j_{RF}} |m\ket \binom{2j_{RF}}{j_{RF} + m}^{1/2}\cos^{j_{RF}+m}(\frac{\tet}{2}) \sin^{j_{RF}-m}(\frac{\tet}{2})e^{-i m
    \fii}.
$ The rotated POVM is given by  $\hat{P}^{j_{RF}}_{j_S \, m}(\al) =
\bra \al | \hat{\Pi}_{j_{RF}+m} |\al \ket$ and the corresponding
parity operator by $\hat{P}^{j_{RF}}_{j_S}(\al)=
\sum^{j_S}_{m=-j_S}(-1)^{j_S-m} \hat{P}^{j_{RF}}_{j_S\,m}(\al)$.
Finally, the RF dependent correlation function reads
$E_{j_{RF}}(\al,\be) = \bra\Psi_{j_S}^-|\hat{P}^{j_{RF}}_{j_S}(\al)\otimes
\hat{P}^{j_{RF}}_{j_S}(\be)|\Psi_{j_S}^-\ket$.

We consider the situation where all the measurement angles are chosen in
the same plane ($\fii=0$), with the first observer choosing between
settings $\tet_1$ and $\tet_3$ and the second between $\tet_2$ and
$\tet_4$. Taking $\tet_1- \tet_2= \tet_2-\tet_3=\tet_3-\tet_4\equiv
\Delta \tet$, the CHSH inequality reads $S_{j_{RF}}=\left| 3
E_{j_{RF}}(\Delta \tet)-E_{j_{RF}} (3 \Delta \tet) \right| \leq 2$.
For classical reference frames, the angle difference which maximizes
$S$ in the limit of large spin $j_S$ is for $\Delta\tet=\frac{x}{2j_S+1}$,
with $x=1.054$. Note that the angle difference is inversely
proportional to the spin size. In the case of bounded RFs of finite
size $j_{RF}$ it is hard to compute the correlation function
analytically, primarily due to the presence of a large number of non-trivial
Clebsch-Gordan coefficients. As such we evaluate it, and the CHSH expression,
numerically and illustrate the results in Fig. 1. We see that
one needs the size of the RFs to scale at least quadratically with the
size of entangled spins to observe violation.
\begin{figure}
\includegraphics[width=8cm]{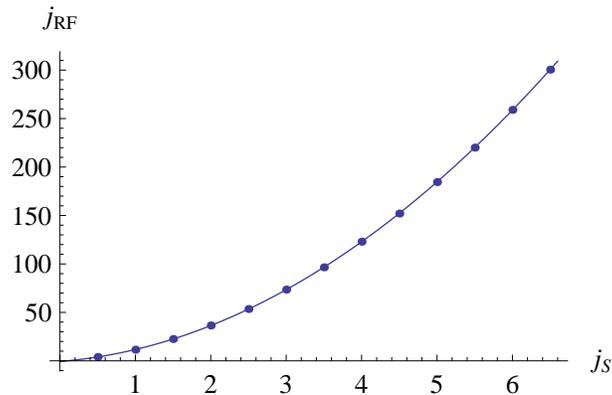}

\caption{Minimal size of (spin) reference frame $j_{RF}$ needed for
violation of the Bell inequality for measurements on a pair of
entangled spins of size $j_S$. The dots are calculated numerically,
the line is the extrapolated fit $j_{RF} \simeq 6 j_S^2 + 6 j_S$
(see text).}
\end{figure}

We now give a heuristic argument to support our numerical findings.
Consider a coherent state of length $j_{RF}$ pointing in a direction
$(\tet,\fii=0)$ and a measurement of spin projection along the
$z$-axis. The probability to obtain outcome $m$ for the spin $z$
component obeys a binomial distribution $p(m)=
\binom{2j_{RF}}{j_{RF} + m}q^{j_{RF}+m} (1-q)^{j_{RF}-m}$, where
$q=\cos^2(\frac{\tet}{2})$. For large $j_{RF}$ this is approximately
a Gaussian centered in $j_{RF} \cos\tet$ and with variance $\sigma^2
= \frac{1}{2} j_{RF} \sin(\tet) \sim j_{RF}$. It can be visualized
as an arrow pointing toward $\tet$ with an angle uncertainty
$\frac{1}{\sigma} \sim \frac{1}{\sqrt{2j_{RF}}}$. Using this as a RF
it is impossible to distinguish between directions at angles closer
than this amount. On the other hand, violation of Bell's inequalities
requires us to measure setting directions at angles that differ at
the order of $\Delta\tet \sim \frac{1}{j_S}$. To achieve this
precision one needs $\frac{1}{\sigma} < \Delta\tet$, which gives the
heuristic bound $j_{RF}
> j{_S}{^2}$. Fitting our numerical results with a quadratic law we indeed find
the formula $j_{RF} \simeq 6 j_S^2 + 6 j_S$ (Fig. 1).
(For higher order fits we obtain coefficients close to zero for the powers higher than two).

In conclusion, we have shown how a bounded RF limits the ability of
entangled systems to exhibit genuine quantum features, as characterized
by violation of Bell's inequalities. This can be relevant in
situations where, to implement quantum information tasks, only
relational degrees of freedom can be exploited (for example, when the quantum channel is subject to a global noise). We focused on the restrictions derived from the lack of a directional RF, considering other restrictions would impose additional requirements on the resources needed (see, e.g., \cite{ashhab}). 

Quantum behavior is generally not observed in macroscopic systems. One reason is that an extremely high experimental resolution would be required to observe quantum phenomena, larger than what can be practically reached [7]. The quantum nature of any physical (i.e. finite) reference frame employed in an experiment gives a fundamental limitation on the maximally achievable resolution. Our results set a lower bound on the resources (i.e. the strength of quantum reference frames) needed to obtain the resolution that is necessary for observation of quantum features of a system with a given size. 
For example, a small iron magnet can have a magnetic field of around 100 $G$. Suppose that an entangled state of a pair of spins  each with size $j_S\simeq \frac{\hbar 100\,G}{\mu_B}
\simeq 10^{21}$ ($\mu_B$ is the Bohr magneton) were available and could be sufficiently protected from decoherence. Even in this case,
according to our analysis, no violation of local realism is possible unless the RFs correspond to magnetic fields at least of order $10^4$ $G$. This is much larger than what can be generally found in nature (but still not impossible to produce).

Typically, quantum coherence in a system is quickly lost due to its interaction with the environment, but the coherence is preserved in correlations between the system and the environment \cite{zurek}. A possible explanation of why we observe classicality of the macroscopic world rather than these quantum correlations is that they  have no operational meaning unless a sufficiently strong RF is available (cf, e.g., \cite{gambini2}). But in everyday experience such RFs are not available for systems of the size of the environment. Our analysis suggests that, for directional RFs, at least a quadratic scaling with the size of the system would be required to demonstrate the existence of the quantum correlations.

We acknowledge many useful conversations with Steve Bartlett, Johannes Kofler, Yeong-Cherng Liang, Tomasz Paterek and Robert Spekkens. This work was supported by the Austrian Science Foundation FWF within Projects No. P19570-N16, SFB and CoQuS No. W1210-N16,  the
European Commission, Project QAP (No. 015848) and the UK Engineering and Physical Sciences Research Council.


\begin{thebibliography}{99}
\bibitem{note}To physically implement measurements of total spin, one must be able to entangle the two particles. In practice, information about the alignment of the beams containing the particles with respect to the lab would need to be available. However, typically this information about beam alignment does not contain any information about the particle spins - i.e. the internal degrees of freedom of the two particles. Therefore, one can lack RF for internal degrees of freedom and still be able to perform the required measurements in the lab.
\end{thebibliography}
\end{document}